\documentclass[12pt]{article}

\usepackage{amsfonts}
\usepackage[centertags]{amsmath}
\usepackage{amssymb}
\usepackage{cite}
\usepackage{psfrag}
\usepackage{graphicx}
\usepackage{bbm,bm}
\usepackage{epsfig, multicol}

\allowdisplaybreaks
\begin{document}

\topmargin 0pt
\oddsidemargin 0mm
\def\be{\begin{equation}}
\def\ee{\end{equation}}
\def\bea{\begin{eqnarray}}
\def\eea{\end{eqnarray}}
\def\ba{\begin{array}}
\def\ea{\end{array}}
\def\ben{\begin{enumerate}}
\def\een{\end{enumerate}}
\def\nab{\bigtriangledown}
\def\tpi{\tilde\Phi}
\def\nnu{\nonumber}
\newcommand{\eqn}[1]{(\ref{#1})}

%\textheight=9in  \textwidth=6.5in 
%\headheight=0in   \headsep=0in
%\topmargin=0.2in    \oddsidemargin=0.0in
%\renewcommand{\baselinestretch}{1.1}
%\parskip .2m
%\parindent .5cm
%\jot=5pt   %ver. Spacing in eqnarray
%\arraycolsep=2pt    %hor. spacing in eqnarray
\newcommand{\half}{{\frac{1}{2}}}
\newcommand{\vs}[1]{\vspace{#1 mm}}
\newcommand{\dsl}{\pa \kern-0.5em /} 
\def\a{\alpha}
\def\b{\beta}
\def\g{\gamma}\def\G{\Gamma}
\def\d{\delta}\def\D{\Delta}
\def\ep{\epsilon}
\def\et{\eta}
\def\z{\zeta}
\def\t{\theta}\def\T{\Theta}
\def\l{\lambda}\def\L{\Lambda}
\def\m{\mu}
\def\f{\phi}\def\F{\Phi}
\def\n{\nu}
\def\p{\psi}\def\P{\Psi}
\def\r{\rho}
\def\s{\sigma}\def\S{\Sigma}
\def\ta{\tau}
\def\x{\chi}
\def\o{\omega}\def\O{\Omega}
\def\k{\kappa}
\def\pa {\partial}
\def\ov{\over}
\def\nn{\nonumber\\}
\def\ud{\underline}
%%%%%%%%%%%%%%%%%%%%%%%%%%%%%%
\begin{flushright}
%UM-TH-00-16\\
%SINP/TNP/00-20\\
%arXiv:YYMM.NNNN\\
%  
\end{flushright}
%\preprint{}
\begin{center}
{\large{\bf Decoupling of gravity on non-susy D$p$ branes}}

\vs{10}

{Kuntal Nayek\footnote{E-mail: kuntal.nayek@saha.ac.in} and Shibaji Roy\footnote{E-mail: shibaji.roy@saha.ac.in}}

\vs{4}

{\it Saha Institute of Nuclear Physics\\
1/AF Bidhannagar, Calcutta 700064, India\\}

\end{center}

\vs{15}

\begin{abstract}
We study the graviton scattering in the background of non-susy D$p$ branes of type II string theories consisting 
of a metric, a dilaton
and a $(p+1)$ form gauge field. We show numerically that in these backgrounds graviton experiences a scattering potential 
which takes the form of an infinite barrier in the low energy (near brane) limit for $p\leq 5$ and therefore 
is never able to reach the branes. This shows, contrary to what is known in the literature, that gravity indeed decouples 
from the non-susy D$p$ branes for $p \leq 5$. For non-susy D6 brane, gravity couples as there is no such barrier for 
the potential. To give further credence to our claim we solve the scattering
equation in some situation analytically and calculate the graviton absorption cross-sections on the non-susy branes and 
show that they vanish for $p \leq 4$ in the low energy limit.
This shows, as in the case of BPS branes, that gravity does decouple for non-susy D$p$ branes for $p\leq 4$ but it
does not decouple for D6 brane as the potential here is always attractive. We argue for the non-susy D5 brane 
that depending on one of the parameters of the solution gravity either always decouples (unlike the BPS D5 brane) or it 
decouples when the energy of the graviton is below certain critical value, otherwise it couples, very similar to BPS D5 brane.      
\end{abstract}

\newpage

\noindent{\it 1. Introduction} : The AdS/CFT correspondence 
\cite{Maldacena:1997re, Witten:1998qj, Gubser:1998bc, Aharony:1999ti} is a holographic equivalence
between certain quantum field theory without gravity and supergravity or string theory on a specific curved background.
This was first conjectured by Maldacena \cite{Maldacena:1997re} in the context of BPS D3-brane of type IIB 
string theory. The curved background
in this case is the five-dimensional AdS space (apart from a 5-sphere) and the non-gravitational quantum field 
theory is the four-dimensional ${\cal N}=4$ super Yang-Mills theory which is a conformal field theory living on the 
boundary of AdS$_5$. Thus the equivalence is holographic and it is also a strong-weak duality symmetry which is very
useful to extract information about strongly coupled field theory from weakly coupled string (or supergravity) theory 
and vice-versa. String theory in the background of D3 brane consists of three parts: (i) theory living on the brane, (ii)
theory living in the bulk and (iii) interaction between the two. It has been shown that there exists a decoupling 
(or a low energy) limit by which the interaction can be made to vanish and thus the field theory living on the brane gets
decoupled from the bulk gravity theory. More precise way to see that the gravity decouples from the brane is to study
the graviton scattering in the background of D3-brane \cite{Klebanov:1997kc,Gubser:1997yh,Gubser:1998iu}. Both from studying 
the scattering potential and also from the calculation of graviton absorption cross-section 
\cite{Das:1996wn, Das:1996jy, Das:1996we, Klebanov:1997kc, Gubser:1997yh,Gubser:1998iu} 
one can see, that the scattering potential takes the form of an infinite 
barrier and the absorption cross-section vanishes in the decoupling limit. Therefore, a graviton propagating in the bulk
and approaching the brane is never able to reach the brane in the decoupling limit and thus the bulk gravity gets decoupled
from the brane.         

The correspondence holds good even when there are less number of supersymmetries and with \cite{Klebanov:1998hh} 
or without conformal symmetries \cite{Klebanov:2000nc, Klebanov:2000hb} .
So, for example, BPS D$p$ branes (for $p\leq 5$ and $\neq 3$) of type II string theories has a decoupling limit in which the 
field theories on the brane get decoupled from the bulk gravity \cite{Itzhaki:1998dd}. The field theories in these 
cases have 16 supercharges
and no conformal symmetries giving rise to a non-AdS/non-CFT correspondence. One can explicitly check that decoupling
of gravity on the brane indeed occurs by studying the graviton scattering on these branes. As in BPS D3 brane case, here
also one can see that the graviton equation of motion takes the form of a Schr\" odinger-like equation, where the potential
becomes infinite in the decoupling limit, indicating that gravity gets decoupled from the brane. One can further check the
graviton absorption cross-section on the brane and indeed one finds that it vanishes in the decoupling limit which clearly
indicates that the graviton does not reach the brane and the bulk gravity gets decoupled from the brane 
world-volume theory \cite{Alishahiha:2000qf}.
Decoupling of gravity does not occur for BPS D6 brane as the scattering potential in this case does not have a barrier. 
BPS D5 brane is a border-line between
D6 brane and other D$p$ branes. Here, the scattering potential indicates that gravity decouples if the energy carried by
the graviton is below certain critical value, and above that value gravity couples to BPS D5 brane \cite{Alishahiha:2000qf}.

It is generally believed that AdS/CFT-like correspondence must hold good for more general backgrounds and even for the 
non-supersymmetric backgrounds. However, there is no explicit calculation in the literature showing the decoupling of gravity 
for the non-supersymmetric gravitational systems\footnote{Some earlier comments and calculation about the
non-decoupling of gravity for non-susy branes can be found in \cite{Brax:2000cf} and \cite{Brax:2001dc}, however, we differ 
with their conclusion.}. 
In this paper, we study the graviton scattering on the known non-supersymmetric 
(non-susy) D$p$-brane solutions of type II string theories \cite{Zhou:1999nm, Lu:2004ms}. As a simple exercise we first look
at the dynamics of a scalar coupled minimally to this background. We write the equation of motion in the string frame and show
that the scalar satisfies, in this background, a Schr\" odinger-like equation with certain potential. The purpose of studying 
the minimally coupled scalar
is that when we study the graviton scattering next, we will see that it will essentially (upto a multiplicative function) satisfy 
the same Schr\" odinger-like equation with the same potential in the background of non-susy D$p$ brane solutions.  
We study the scattering potential 
numerically and show that in the decoupling limit the potentials act like an infinite barriers for the graviton to reach the 
brane for all $p\leq 5$. However, for $p=6$, there is no barrier for the potential and it always goes to negative infinity 
indicating that the gravity in this case does not decouple from the brane. To further strengthen our claim we compute the 
graviton absorption cross-sections for the non-susy
D$p$ branes. For this we need to solve the Schr\" odinger-like scattering equation. It is in general difficult to solve
the equation, but, the equation can be solved both at the far region and at the near region and this is what is needed for
obtaining the expression for the absorption cross-sections. However, unlike the BPS brane case 
\cite{Klebanov:1997kc, Gubser:1997yh, Alishahiha:2000qf} the near region solution for
the non-susy case can be obtained only if the parameters of the solutions satisfy certain conditions discussed in section 4. Then 
the solutions in the far region and the near region can be matched
in the overlapping region and that fixes the arbitrary constant in the solution\footnote{Similar calculation has been done for
BPS D3 brane in \cite{Klebanov:1997kc}. See also \cite{Das:1996wn, Das:1996jy, Das:1996we} for some earlier calculation of similar 
type for D1/D5 black holes.}. From the form of the solution, we can
straightforwardly obtain the expressions for the graviton absorption cross-sections on the brane. Then we show that the 
cross-sections vanish for all non-susy D$p$ branes with $p \leq 4$ in the decoupling limit. We, therefore,
conclude that the bulk gravity indeed decouples for all non-susy D$p$ branes for $p\leq 4$. The calculation
does not work for $p=5$, however, by studying the scattering potential we conclude that here depending on one of the parameters of
the solution the decoupling occurs without any restrictions, otherwise, it occurs only when the energy of the graviton is below 
certain critical value, and above that value it couples. Even for $p=6$, the similar calculation does not go through and so,
we conclude from the scattering potential that the gravity in this case couples as the potential here is attractive.

This paper is organized as follows. In the next section we briefly review the structure of non-susy D$p$ branes and their BPS limits.
Then in section 3, we study the scattering of a minimally coupled scalar and obtain the form of the scattering potential.
In section 4, we study the graviton scattering and study the scattering potential numerically. In section 5, the absorption
cross-section of the graviton is obtained when the parameters of the solutions satisfy certain conditions. Finally we conclude
in section 6.

\vspace{.5cm}

\noindent{\it 2. Non-susy D$p$ branes and their BPS limits} : In this section we briefly recall the non-susy D$p$ brane solutions
of type II string theories \cite{Lu:2004ms} and see how one can recover the BPS D$p$ brane solutions from them. The type II 
supergravity action we consider is given as,
\be\label{action}
S = \frac{1}{16\pi G_{10}}\int d^{10}x \sqrt{-g}\left[e^{-2\phi}\left(R + 4\partial_\mu \phi \partial^\mu \phi\right) - 
\frac{1}{2 \cdot (8-p)!} F_{[8-p]}^2\right]
\ee              
where, $G_{10}$ is the ten dimensional Newton's constant, $g = {\rm det}(g_{\mu\nu})$, with $g_{\mu\nu}$ being the ten dimensional 
string-frame metric 
and $R$ is its curvature scalar, $\phi$ is the dilaton and $F_{[8-p]}$ is an $(8-p)$ RR form-field. 
One can solve the equations of motion following from this action with an appropriate form-field and a static, spherically symmetric 
$p$ brane metric 
(i.e., metric having the isometry ISO($p,\,1$) $\times$ SO($9-p$)) ansatz not respecting the supersymmetry condition and we get the 
non-susy D$p$ brane solutions in the following form \cite{Lu:2004ms}, 
\bea\label{nonsusydp}
& & ds^2 = F(r)^{-\frac{1}{2}}\left(\frac{H(r)}{\tilde{H}(r)}\right)^{\frac{\delta}{2}}\left(
-dt^2 + \sum_{i=1}^p(dx^i)^2\right)\nn 
& & \qquad\qquad\qquad\qquad + F(r)^{\frac{1}{2}} \left(\frac{H(r)}{\tilde{H}(r)}\right)^{\frac{\delta}{2}}
\left(H(r) \tilde{H}(r)\right)^{\frac{2}{7-p}}\left(dr^2 + r^2 d\Omega_{8-p}^2\right)\nn
& & e^{2\phi} = F(r)^{\frac{3-p}{2}} \left(\frac{H(r)}{\tilde{H}(r)}\right)^{2\delta}, \qquad F_{[8-p]} = Q {\rm Vol}(\Omega_{8-p})
\eea  
The various functions appearing in the solution are defined as,
\bea\label{functions}
&& H(r) = 1 + \frac{r_p^{7-p}}{r^{7-p}}, \qquad \tilde{H}(r) = 1 - \frac{r_p^{7-p}}{r^{7-p}}\nn
&& F(r) = \left(\frac{H(r)}{\tilde{H}(r)}\right)^\alpha \cosh^2\theta - \left(\frac{\tilde{H}(r)}{H(r)}\right)^\beta \sinh^2\theta
\eea
In the above we have suppressed the string coupling constant $g_s$ which is assumed to be small, $Q$ is the RR charge and 
${\rm Vol}(\Omega_{8-p})$ is the volume form
of the transverse $(8-p)$-dimensional unit sphere. Also, $\alpha$, $\beta$, $\delta$, $\theta$, $Q$ and $r_p$ ($r_p$ has the
dimension of length, but its actual form is different for different $p$ branes) are six parameters
characterizing the solution. However, not all of them are independent. In fact, there are three relations among them following from
the consistency of the equations of motion and they are,
\bea\label{relations}
&& \alpha - \beta = \frac{p-3}{2} \delta\nn
&& \frac{1}{2} \delta^2 + \frac{1}{2}\alpha\left(\alpha-\frac{p-3}{2}\delta\right) = \frac{8-p}{7-p}\nn
&& Q = (7-p) r_p^{7-p}(\alpha+\beta)\sinh2\theta
\eea
So, we can use these relations to eliminate three of the above six parameters and therefore, we take $\delta$, $\theta$ and $r_p$ as
the independent parameters\footnote{Attempts have been made to give physical meaning to these three parameters in terms of number of
branes, number of anti-branes and a tachyon vev or tachyon parameter in \cite{Brax:2000cf, Lu:2004dp}.} of the solution. In fact one can 
obtain $\alpha$, $\beta$ in terms of $\delta$ from the second relation
in \eqref{relations} as,
\bea\label{alphabeta} 
&& \alpha = \sqrt{\frac{2(8-p)}{7-p} - \frac{(p+1)(7-p)}{16}\delta^2} + \frac{p-3}{4}\delta\nn
&& \beta = \sqrt{\frac{2(8-p)}{7-p} - \frac{(p+1)(7-p)}{16}\delta^2} - \frac{p-3}{4}\delta
\eea
where for the reality of the parameters $\delta$ must be constrained as,
\be\label{delta}
|\delta| \leq \frac{4}{7-p}\sqrt{\frac{2(8-p)}{p+1}}
\ee 
One may think that there are too many parameters in the solution which may violate Birkhoff's
theorem. However, note that because of the form of $\tilde{H}(r)$ given in \eqref{functions}, the solution has a singularity at $r=r_p$,
and therefore, Birkhoff's theorem does not apply here. So, $r>r_p$ is the physical region. We have also taken $(\alpha+\beta)$, $\theta$ and
$Q$ to be positive semi-definite without any loss of generality. Note that the non-susy D$p$ brane solutions given in \eqref{nonsusydp} is
asymptotically flat and is magnetically charged. The corresponding electrically charged solutions can be obtained by first writing the metric
in the Einstein frame using $g_{\mu\nu}^{\rm E} = e^{-\frac{\phi}{2}} g_{\mu\nu}$ and then making use of the transformation
$g_{\mu\nu}^{\rm E} \to g_{\mu\nu}^{\rm E}$, $\phi \to -\phi$ and $F \to e^{\frac{(3-p)\phi}{2}} \ast F$, where $\ast$ denotes the Hodge dual and finally
going back to the string frame by the reverse metric transformation just given. For the 
electrically charged solution the gauge field takes the form,
\be\label{gauge}
A_{[p+1]} = \frac{1}{2}\sinh2\theta\,\left(\frac{C(r)}{F(r)}\right) dt\wedge \cdots \wedge dx^p, \quad {\rm with,}\,\, C(r) = 
\left(\frac{H(r)}{\tilde{H}(r)}\right)^\alpha - \left(\frac{\tilde{H}(r)}{H(r)}\right)^\beta
\ee       
and $F(r)$ as given in \eqref{functions}. We will, however, work with the magnetically charged solutions. 

Now to obtain the BPS D$p$ brane from the non-susy D$p$ brane solutions \cite{Lu:2004ms} given in \eqref{nonsusydp}, we notice that if we
take $r_p \to 0$, then both $H(r)$ and $\tilde{H}(r)$ go to 1, but the function
\be\label{Fgoesto}
F(r) \to 1 + 2(\alpha \cosh^2\theta + \beta \sinh^2\theta) \frac{r_p^{7-p}}{r^{7-p}}
\ee
Now if we further take $\theta \to \infty$ such that the product $2(\alpha+\beta)\sinh^2\theta r_p^{7-p} = \bar{r}_p^{7-p} = {\rm fixed}$,
then $F(r) \to \bar{H}(r)$, where $\bar{H}(r) = 1 + \bar{r}_p^{7-p}/r^{7-p}$ is the standard Harmonic function and the solution 
\eqref{nonsusydp} reduces exactly to the standard BPS D$p$ brane solutions. Note that now all the extra paramaters from the non-susy
D$p$ brane solutions disappear and the BPS solutions are characterized by a single parameter $\bar{r}_p$ as expected.

In order to study the graviton scattering we first rewrite it in a more convenient form by introducing a new radial coordinate given by,
\be\label{newcoord}
r = \rho \left(\frac{1+\sqrt{G(\rho)}}{2}\right)^{\frac{2}{7-p}}, \qquad {\rm where,}\,\, G(\rho) = 1 + \frac{4r_p^{7-p}}{\rho^{7-p}} \equiv 
1 + \frac{\rho_p^{7-p}}{\rho^{7-p}}
\ee
In this new coordinate we have
\bea\label{coordchange}
&& H = \frac{2\sqrt{G(\rho)}}{1+\sqrt{G(\rho)}}, \qquad \tilde{H} = \frac{2}{1+\sqrt{G(\rho)}}\nn
&& dr^2 + r^2 d\Omega_{8-p}^2 = \left(\frac{1+\sqrt{G(\rho)}}{2}\right)^{\frac{4}{7-p}}\left[\frac{d\rho^2}{G(\rho)} + d\Omega_{8-p}^2\right]
\eea
Using \eqref{coordchange}, we can rewrite the non-susy D$p$ brane solution \eqref{nonsusydp} as,
\bea\label{nonsusydpnew}
& & ds^2 = F(\rho)^{-\frac{1}{2}} G(\rho)^{\frac{\delta}{4}}\left(
-dt^2 + \sum_{i=1}^p(dx^i)^2\right) + F(\rho)^{\frac{1}{2}} G(\rho)^{\frac{1}{7-p} + \frac{\delta}{4}}\left(\frac{d\rho^2}{G(\rho)} + 
\rho^2 d\Omega_{8-p}^2\right)\nn
& & e^{2\phi} = F(\rho)^{\frac{3-p}{2}} G(\rho)^{\delta}, \qquad F_{[8-p]} = Q {\rm Vol}(\Omega_{8-p})
\eea 
where $G(\rho)$ is as given in \eqref{newcoord} and 
\be\label{frho}
F(\rho) = G(\rho)^{\frac{\alpha}{2}} \cosh^2\theta - G(\rho)^{-\frac{\beta}{2}} \sinh^2\theta
\ee
The parameter relations remain exactly the same as given in \eqref{relations}. The BPS limit now would be given as $\rho_p \to 0$,
$\theta \to \infty$, such that $(1/2)(\alpha+\beta)\rho_p^{7-p} \sinh^2\theta = \bar{\rho}_p^{7-p} = {\rm fixed}$ (where 
$\bar{\rho}_p^{7-p} = 4\bar{r}_p^{7-p}$).
Note that in this case $G(\rho) \to 1$ and $F(\rho) \to \bar{H}(\rho)$, where $\bar{H}$ is the standard harmonic function of a BPS
D$p$ brane and thus \eqref{nonsusydpnew} reduces precisely to the BPS D$p$ brane solution. We will use the solution \eqref{nonsusydpnew}
to study both the minmally coupled scalar scattering and the graviton scattering in the following. 

\vspace{.5cm}

\noindent{\it 3. Minimally coupled scalar scattering} : In this section we will study the scattering of a massless scalar ($\varphi$) 
coupled minimally to the non-susy D$p$ brane background and obtain the form of the scattering potential it experiences while moving in the
background. The reason for studying this, as we will see, is that the graviton also experiences 
the same scattering potential (studied in the next section) as the scalar. The relevant part of the action in this case is
\be\label{scalaraction}
S_{\rm scalar}=\frac{1}{4\pi G_{10}}\int d^{10}x\sqrt{-g}e^{-2\phi}\partial_\mu\varphi\partial^\mu\varphi
\ee 
where $g_{\mu\nu}$ is the background metric in string frame and $\phi$ is the dilaton field. Note that we have omitted the Ricci 
scalar, the kinetic energy of the dilaton and the RR form-field terms from the action since they don't contribute to the scalar equation of
motion. From \eqref{scalaraction} the scalar $\varphi$ is found to satisfy the following equation of motion,
\be\label{scalareom}
D_\mu \partial^\mu \varphi - 2 D_\mu\phi\partial^\mu \varphi = 0
\ee
Now we assume that $\varphi$ has spherical symmetry in the transverse space (that means we are considering only scalar s-wave, since
the higher partial waves will give an additional repulsive centrifugal term and will have greater chances of decoupling) and is independent 
of the spatial coordinates of the brane (assumed for simplicity). Thus $\varphi$ has the form,
\be\label{varphi}
\varphi = \Phi(\rho) e^{i\omega t}
\ee
Now using \eqref{varphi}, the scalar equation of motion \eqref{scalareom} with the background given in \eqref{nonsusydpnew} 
reduces to the following
second order differential equation,
\be\label{2ndordereq}
\partial_\rho^2 \Phi(\rho) + \left[\frac{8-p}{\rho} + \frac{\partial_\rho G(\rho)}{G(\rho)}\right]\partial_\rho \Phi(\rho) + \omega^2
F(\rho) G(\rho)^{-\frac{6-p}{7-p}}\,\Phi(\rho) = 0
\ee
Redefining the radial coordinate as, $u = \omega \rho$ and introducing a function $g(u)$ by 
\be\label{newfn1}
\Phi(u) = k(u) g(u), \qquad {\rm where,} \quad k(u) = \frac{1}{\sqrt{u^{8-p} G(u)}},
\ee
\eqref{2ndordereq} takes the form of a Schr\" odinger-like equation in terms of the new function as,
\be\label{scheq1}
\left(\partial_u^2 - V(u)\right) g(u) = 0
\ee
where,
\be\label{potential}
V(u) = \frac{(8-p)(6-p)}{4u^2} - \frac{1}{4} \left(\frac{\partial_u G(u)}{G(u)}\right)^2 - F(u) G(u)^{-\frac{6-p}{7-p}}
\ee
and
\be\label{gufu}
G(u) = 1 + \frac{(\omega \rho_p)^{7-p}}{u^{7-p}}, \qquad F(u) = G(u)^{\frac{\a}{2}} \cosh^2\theta - G(u)^{-\frac{\b}{2}} \sinh^2\theta.
\ee
Once we have the form of $g(u)$ by solving the equation \eqref{scheq1}, we can obtain the form of the scalar as 
\be\label{dilatonsoln}
\varphi(t,\rho) = 
\frac{1}{(\omega \rho)^{\frac{8-p}{2}}} \frac{g(\omega\rho)}{\sqrt{G(\omega\rho)}} e^{i\omega t}.
\ee
Eq.\eqref{potential} represents the scattering potentials the minimally coupled scalar experiences while moving in the background of 
non-susy D$p$ branes. We will analyze these potentials \eqref{potential} in the next section.

\vspace{.5cm}

\noindent{\it 4. Graviton scattering: Potentials} : In this section we will study the graviton scattering on the non-susy D$p$ branes 
given in section 2. We will obtain the form of the scattering potential (which will have the same form \eqref{potential} as 
the scattering potential of a minimally coupled scalar obtained in the previous section) and
analyze it numerically. To study the scattering we need the linearized equations of motion from the
action \eqref{action} in the background \eqref{nonsusydpnew}. The equations of motion can be 
linearized by perturbing the background metric as, $g_{\mu\nu} = \bar{g}_{\mu\nu} + h_{\mu\nu}$, where background metric
is denoted with a `bar'.
The linearized forms of the dilaton and the graviton equations 
of motion are \cite{Alishahiha:2000qf},
\bea
& & \Gamma(h)^\mu_{\mu\nu}\partial^\nu\phi + h^{\mu\nu}D_\mu\partial_\nu\phi = - \frac{1}{4} \bar{R}_{\mu\nu}h^{\mu\nu} + \frac{1}{4} 
\left(D_\mu D_\nu h^{\mu\nu} - D^2 h_\mu^\mu\right) + h^{\mu\nu}\partial_\mu\phi\partial_\nu\phi\label{dilatoneom}\\
& & D_{(\mu} D_\rho h^\rho_{\nu)} - \frac{1}{2} D^2 h_{\mu\nu} - \frac{1}{2} D_\mu D_\nu h_\rho^\rho + \bar{R}_{\rho(\mu} h_{\nu)}^{\,\rho} + 
\bar{R}_{\nu\rho\sigma\mu} h^{\rho\sigma}\nn
& & = 2\Gamma(h)_{\mu\nu}^\rho\partial_\rho\phi - \frac{e^{2\phi}}{2 \cdot (8-p)!}\left((8-p)(7-p) h^{\rho\sigma} 
F_{(\mu|\rho}^{\,\,\,\,\,\,\,\,\,\mu_1 \cdots \mu_{6-p}} F_{\nu)\sigma\mu_1 \cdots \mu_{6-p}}\right.\nn
& & \qquad\qquad\qquad \left. - \frac{8-p}{2} \bar{g}_{\mu\nu} h^{\rho\sigma} F_{\rho}^{\,\,\,\mu_1 \cdots \mu_{7-p}} F_{\sigma\mu_1 \cdots \mu_{7-p}} 
+ \frac{1}{2} h_{\mu\nu} F^2\right)\label{gravitoneom} 
\eea
Again, for the reasons given before for the scalar, we here consider only the graviton s-wave and assume (for simplicity) it to be independent 
of the spatial directions of the brane. So, the graviton takes the form,
\be\label{grav}
h_{\mu\nu} = \epsilon_{\mu\nu} h(\rho) e^{i\omega t}  
\ee
where $\epsilon_{\mu\nu}$ is the polarization tensor for the graviton. Further, we take gravitons to have the polarizations along the brane only
and therefore, $\epsilon_{\mu\nu} =0$, for $\mu,\nu\,=\, p+1,\,p+2, \ldots, 9$. The transversality conditions of the graviton further restrict 
the polarization tensor, namely, we must have $\epsilon_{\mu 0} = 0$ and for the consistency with \eqref{dilatoneom} we take it to be
traceless, i.e.,  $g^{ab} \epsilon_{ab} = 0$, where 
$a,\,b\,=\, 0,1,\ldots, p$. So, altogether there are $[p(p+1)/2] - 1$ number of possible choices of polarization tensor $\epsilon_{ab}$. Among
them $p(p-1)/2$ are off-diagonal and $p-1$ are diagonal. We choose the only non-zero off-diagonal components as $\epsilon_{12}=\epsilon_{21} = 1$,
and the only non-zero diagonal components as $\epsilon_{11}=-\epsilon_{22}=1$ satisfying all the restrictions on the polarization tensor. Since
both types of polarizations give the same equation for $h(\rho)$, we give its form using \eqref{gravitoneom} and \eqref{nonsusydpnew} as,
\bea\label{gravitoneq}
& & \partial_\rho^2 h(\rho) + \left[\frac{8-p}{\rho} + \frac{\partial_\rho F}{F} + \left(1-\frac{\delta}{2}\right)
\frac{\partial_\rho G}{G}\right]\partial_\rho h(\rho)\nn
& & +\left[\omega^2 F G^{-\frac{6-p}{7-p}} + \frac{1}{4}\left(\frac{\partial_\rho F}{F} - 
\frac{\delta}{2}\frac{\partial_\rho G}{G}\right)^2
-\frac{1}{2} \left(\frac{Q}{\rho^{8-p}}\right)^2 F^{-2} G^{-\frac{3-p}{4}\delta-2}\right]h(\rho) = 0
\eea
Again redefining the radial coordinate by $u = \omega \rho$, and introducing the same function $g(u)$ as before, but now is related 
to $h(u)$ by 
\be\label{newfunction}
h(u) = \tilde{k}(u) g(u), \qquad {\rm where,} \quad \tilde{k}(u) = \frac{1}{\sqrt{u^{8-p} F(u) G(u)^{1-\frac{\delta}{2}}}},
\ee
we find, after some manipulation, that \eqref{gravitoneq} reduces precisely to the same Schr\" odinger-like form in terms of $g(u)$ 
as \eqref{scheq1} with the potential as given in
\eqref{potential}. However, the graviton solution $h_{\mu\nu}(t, \rho)$, is obviously different (as can be seen from \eqref{newfunction})
from that of the minimally coupled scalar \eqref{dilatonsoln}. It is therefore clear that the graviton also experiences the same 
scattering potentials 
as a minimally coupled scalar when it moves in the background of non-susy D$p$ branes.

\begin{figure}[ht]
  \begin{center}
    \includegraphics[width=0.75\textwidth,height=7.5cm]{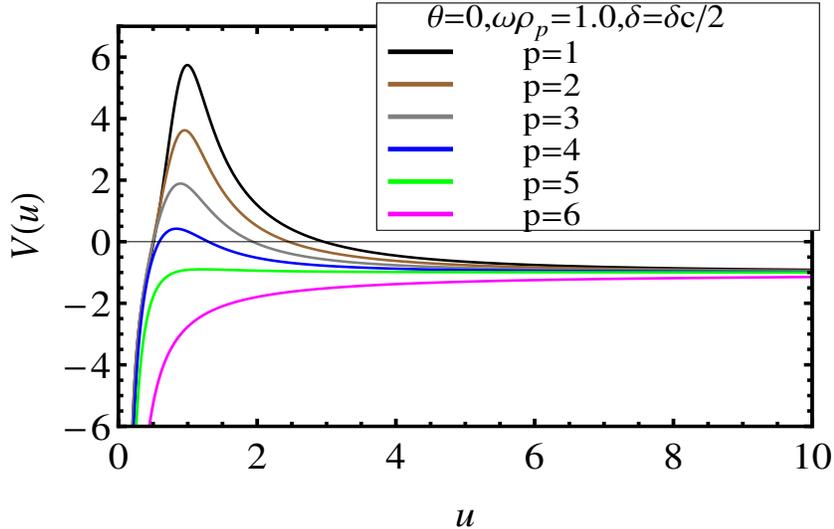}
    \caption{\it Plot of the potential $ V(u) $ given in \eqref{potential} vs $ u $. Here the potential is given for $\theta=0 $ , 
$\omega\rho_p=1.0$ and 
$\delta=\delta_c/2=\frac{2}{7-p}\sqrt{\frac{2(8-p)}{p+1}}$.}
    \label{fig:or10}
  \end{center}
\end{figure}

\begin{figure}[ht]
  \begin{center}
    \includegraphics[width=0.75\textwidth,height=7.5cm]{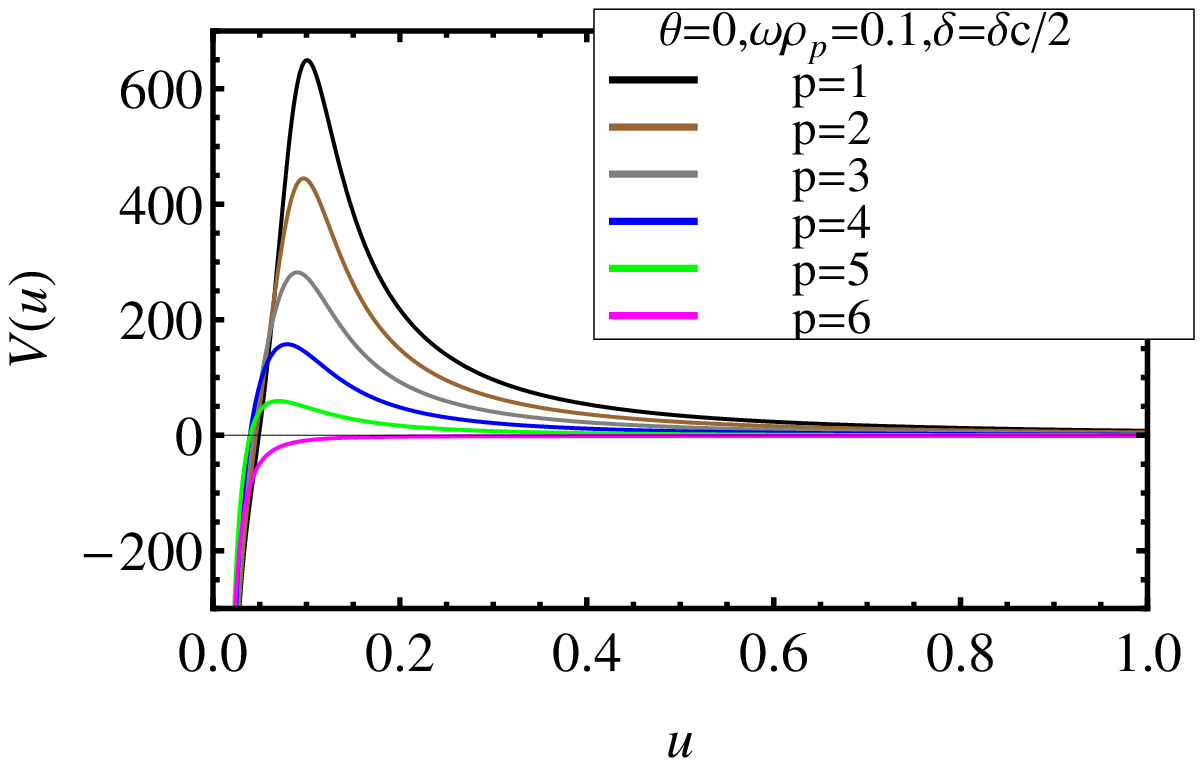}
    \caption{\it Plot of the potential $ V(u) $ given in \eqref{potential} vs $ u $. Here the potential is given for $\theta=0$, 
$\omega\rho_p=0.1$ and 
$\delta=\delta_c/2=\frac{2}{7-p}\sqrt{\frac{2(8-p)}{p+1}}$.}
    \label{fig:or01}
  \end{center}
\end{figure}

\begin{figure}[ht]
  \begin{center}
    \includegraphics[width=0.75\textwidth,height=7.5cm]{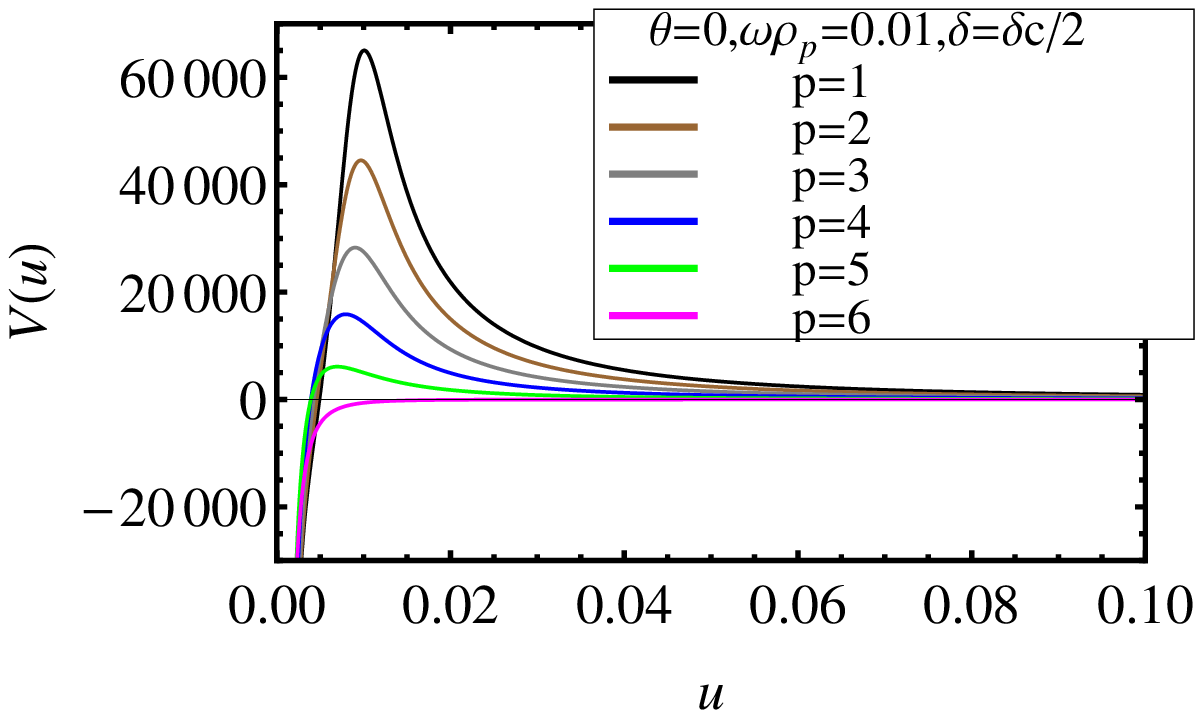}
    \caption{\it Plot of the potential $ V(u) $ given in \eqref{potential} vs $ u $. Here the potential is given for $\theta=0$, 
$\omega\rho_p=0.01$ and 
$\delta=\delta_c/2=\frac{2}{7-p}\sqrt{\frac{2(8-p)}{p+1}}$.}
    \label{fig:or001}
  \end{center}
\end{figure}

We now take a close look at the graviton scattering potential $V(u)$ given in \eqref{potential}. First we note that it has three
terms of which the first one is positive for $p \leq 5$ and is zero for $p=6$, and the second and third terms are always negative. So,
the potential has both repulsive and attractive pieces for $p \leq 5$, but it is always (for all $u$) attractive for $p=6$. In fact 
this is the reason that gravity does not decouple from the non-susy D6 branes. However, for all $p$, when $u \to \infty$, i.e., in the far
region $G(u) = 1+ (\omega \rho_p)^{7-p}/u^{7-p} \to 1$ and as a result $F(u) = G(u)^{\alpha/2} \cosh^2\theta - G(u)^{-\beta/2} \sinh^2\theta \to 1$ 
and so, we find from \eqref{potential} that the potential $V(u) \to -1$. On the other hand in the near region, i.e., when $u \to 0$,
$V(u) \to -1/(4u^2) - F(u) G(u)^{-(6-p)/(7-p)}$ and since $F(u)$ is positive and the second term is negative, $V(u) \to -\infty$. So, one
might think that since in the near region the potential is attractive, gravity will not decouple from the brane. But this is not true. The
point is that in between $u=\infty$ and $u=0$, there might exist some $u$, where the potential can have a maximum. We find that this is
indeed the case, however, it is difficult to find the value of $u$ analytically where the maximum of the potential occurs because of its
complicated form. So, we have 
plotted the potentials $V(u)$ given in \eqref{potential} versus $u$ for different values of the parameters in Figs. 1, 2, and 3. We have
taken $\theta=0$\footnote{This is taken for simplicity. However, note that by the relation \eqref{relations}, this implies that the brane
is chargeless and this is a characteristic of a non-susy brane as a BPS brane should always be charged. Therefore, when we vary $\omega\rho_p$
in the plots Figs 1, 2, 3, the branes always remain non-susy even when we take the decoupling limit, namely, $\omega\rho_p \to 0$.} 
and $\delta = \delta_c/2$, where $\delta_c = \frac{4}{7-p}\sqrt{\frac{2(8-p)}{p+1}}$ which is the maximum allowed value of
$|\delta|$. These are some typical values of the parameters to see the variation of the potential $V(u)$. Actually we have seen that the 
variations of these two parameters have little effects on the potential. In Figs. 1, 2, and 3, we have taken 
$\omega \rho_p = 1,\,0.1,\, {\rm and}\,\,0.01$ and then plotted $V(u)$ for various values of $p$. We notice that for $\omega\rho_p=1$, the
potentials have maxima for each values of $p=1,\,2,\,3,\,4$ (also for $p=5$ but it is not visible because of the scale chosen and
will be clear in Figures 2, 3) except for $p=6$. For $\omega\rho_p = 0.1$, the maxima shift towards the origin and rises
sharply for all values of $p$ except for $p=6$. When $\omega\rho_p = 0.01$, the maxima of the potentials further shift towards the origin 
and take very large values for all values of $p$, but goes to large negative value for $p=6$. Note that as we scale down $\omega \rho_p$
by a factor of 10, the potentials roughly rise by a factor of 100. So, the potentials are approaching to infinity much faster than 
$\omega\rho_p$ going to zero. Thus eventually when we take $\omega\rho_p \to 0$, i.e., in the decoupling limit\footnote{ 
Note that $\rho_p \sim \ell_s$, where $\ell_s$ is the fundamental string length and so, in the decoupling limit $\ell_s \to 0$ implies
$\rho_p \to 0$. $\omega\rho_p$ is the dimensionless parameter which also tends to zero in the decoupling limit.}  the potentials
will rise to positive infinity near the origin for all values of $p$ except for $p=6$. For $p=6$ it will go to negative infinity without
any maximum in between.  
Due to the infinite potential barriers, 
the graviton coming from infinity will not be able to overcome them and reach the brane and therefore bulk gravity gets decoupled on all 
the non-susy D$p$ branes for 
$p \leq 5$ and since for $p=6$, there is no potential barrier, the graviton will reach the brane and therefore gravity does not decouple 
from the non-susy D6 brane. We will discuss more on $p=5$ case towards the end of next section.                    

Now one can easily check that in the BPS limit ($\rho_p \to 0$, $\theta \to \infty$, such that 
$(1/2)(\alpha+\beta)\rho_p^{7-p} \sinh^2\theta = Q/(7-p) = \bar{\rho}_p^{7-p} = {\rm fixed}$), $G(\rho) \to 1$ and 
$F(\rho) \to 1 + Q/\{(7-p)\rho^{7-p}\}$, the standard harmonic function of a BPS D$p$ brane, the potentials $V(u)$ for the non-susy D$p$ branes
as given in \eqref{potential} reduce precisely to the scattering potentials of a graviton (or a minimally coupled scalar)
for a BPS D$p$ branes given in Eq.(4.6) of \cite{Alishahiha:2000qf}.        

To further support our claim that bulk gravity does decouple from the non-susy D$p$ brane world volumes we compute the graviton absorption
cross-sections on the non-susy D$p$ branes in the next section and show that they vanish in the decoupling limit.

\vspace{.5cm}

\noindent{\it 5. Graviton scattering: absorption cross-section} : To compute the graviton absorption cross section we will solve the 
Schr\" odinger-like scattering equation given in \eqref{scheq1},
\be\label{scatteringeq}
\left(\partial_u^2 - V(u)\right) g(u) = 0, \quad {\rm with,}\quad V(u) = \frac{(8-p)(6-p)}{4u^2} - \frac{1}{4} \left(\frac{\partial_u 
G(u)}{G(u)}\right)^2 - F(u) G(u)^{-\frac{6-p}{7-p}}
\ee       
where $G(u)$ and $F(u)$ are as given before in \eqref{gufu}. It is difficult to solve this equation 
in general and so, as usual, we will solve it both in the far region and in the near region and then match the two solutions in the overlapping
region to fix a constant in the solution. In the far region $u \gg \omega\rho_p$ and therefore $G(u) \approx 1$ and $F(u) \approx 1$ and then
\eqref{scatteringeq} reduces to 
\be\label{scatteringfar}
\left[\partial_u^2 + \left(1+\frac{1-4\left(\frac{7-p}{2}\right)^2}{4u^2}\right)\right] g(u) = 0
\ee  
This is the Bessel equation of order $\n = \frac{7-p}{2}$ and so, the solution is,
\be\label{solnfar}
g(u) = c_\infty \sqrt{u} J_{\frac{7-p}{2}}(u)
\ee
where $J(u)$ is the Bessel function and $c_\infty$ is a constant. Now with this solution the graviton takes the form (see \eqref{newfunction})
\be\label{solnfarhu}
h(u) = \frac{g(u)}{\sqrt{u^{8-p} F(u) G(u)^{1-\frac{\delta}{2}}}}  = 
\frac{c_\infty u^{-\frac{7-p}{2}} J_{\frac{7-p}{2}}(u)}{\sqrt{F(u) G(u)^{1-\frac{\delta}{2}}}} 
\ee
Let us define another function by,
\be\label{tildegu}
\tilde{g}(u) = \frac{g(u)}{\sqrt{u^{8-p}}}, \qquad {\rm and\,\,so,}\quad h(u) = \frac{\tilde{g}(u)}{\sqrt{F(u) G(u)^{1-\frac{\delta}{2}}}}
\ee
then the solution in the far region is 
\be\label{solnfar1}
\tilde{g}_\infty(u) = c_\infty u^{-\frac{7-p}{2}} J_{\frac{7-p}{2}}(u)
\ee

Now to find a solution in the near region we note that, unlike in the BPS case \cite{Klebanov:1997kc, Gubser:1997yh, Alishahiha:2000qf}, 
here it is difficult 
to get an analytic solution of the equation \eqref{scatteringeq} in general for all $u$ in the near region. 
So, in the following we go over to certain 
region of space (within the near region) where we can solve \eqref{scatteringeq}
exactly and also have a matching in the overlapping region of the solution \eqref{solnfar1} in the far region and 
that in the near region. This is possible
as we will see if the parameters of the solution are chosen appropriately. 
Now in the near region, we first make a coordinate transformation
\be\label{coordtrans}
z = \frac{2}{5-p} \frac{(\omega\rho_p)^{\frac{7-p}{2}}}{u^{\frac{5-p}{2}}}, \qquad\qquad\qquad\qquad {\rm for,} \quad p<5
\ee
and note that as long as $z \gg (\omega \rho_p)^{\frac{7-p}{2}}$, $u \ll 1$, i.e., we are in the near region where we will find
a solution of Eq.\eqref{scatteringeq}. The solution in the far region ($u \gg \omega\rho_p$) and that in 
the near region ($u \ll 1$) must be matched in the overlapping region to find the arbitrary constant $c_\infty$ and so, $\omega\rho_p \ll 1$
and this gives a restriction on the parameter $\rho_p$ of the non-susy D$p$ brane solutions. 
We further impose the condition that $z \ll \omega\rho_p$. Note that
when $\omega\rho_p \ll 1$, there is no contradiction of this with the previous condition $z \gg (\omega \rho_p)^{\frac{7-p}{2}}$. So, $z$ is in the
range $(\omega \rho_p)^{\frac{7-p}{2}} \ll z \ll \omega\rho_p$.
The scattering equation \eqref{scatteringeq} then can be simplified as,
\bea\label{scatteringnear}
& & \left[\partial_z^2 + \left(\frac{1 - 4\frac{(7-p)^2}{(5-p)^2}}{4z^2}\right) + \left(\frac{\frac{2}{5-p}(\omega\rho_p)}
{z}\right)^{\frac{2(7-p)}{5-p}}\right.\nn
& &\qquad\qquad\qquad\left. \times \left\{1+
\left(b^2-\frac{6-p}{7-p}\right)\left(\frac{z}{\frac{2}{5-p}(\omega\rho_p)}
\right)^{\frac{2(7-p)}{5-p}}\right\}\right]\hat{g}(z)=0
\eea
where we have defined,
\be\label{defs}
b = \sqrt{\frac{\a}{2}\cosh^2\theta + \frac{\b}{2}\sinh^2\theta}, \qquad {\rm and} \qquad \hat{g}(z) = z^{\frac{7-p}{2(5-p)}} g(z)
\ee
Note from \eqref{scatteringnear} that since $z \ll \omega\rho_p$, the first term in the curly bracket will dominate (assuming
$b \sim {\cal O}(1)$ or $\ll 1$) and that will make the differential equation difficult to solve. On the other hand, if we further 
impose the condition on the parameter $b$
such that $z \gg b^{-\frac{5-p}{7-p}} (\omega\rho_p)$, then it is the $b^2$ term in the curly bracket which will dominate. We point out
that this is possible if $b \gg 1$, implying that the parameter $\theta$ is large\footnote{Note that when we plotted the potential in
Figures 1, 2, 3 to show the decoupling we have taken $\theta=0$ for simplicity, but, we have checked that potential has very similar
behavior even for large $\theta$.  This shows that decoupling actually occurs for generic values of
$\theta$, however, a closed form solution of the scattering equation is possible only for large value of $\theta$ satisfying the
condition given above.} and since $\theta$ is an independent parameter of the
solution both the conditions $z \gg (\omega\rho_p)^{\frac{7-p}{2}}$ and the previous one can be simultaneously satisfied. Now with these
conditions \eqref{scatteringnear} can be rewritten in the following form, 
\be\label{scatteringnear1}
\left[\partial_{\hat z}^2 + \left(1 + \frac{1 - 4\frac{(7-p)^2}{(5-p)^2}}{4\hat{z}^2}\right)\right] \hat{g}(\hat{z}) = 0
\ee
Here the radial coordinate $\hat{z}$ is defined as $\hat{z} = b z$. Again we recognize \eqref{scatteringnear1} as the Bessel equation of order 
$\n=\frac{7- p}{5-p}$ and since we are interested in incoming wave for the near region, the relevant solution is,
\be\label{solnnear}
\hat{g}(z) = i \sqrt{b} z^{\frac{1}{2}} \left(J_{\frac{7-p}{5-p}}(bz) + i N_{\frac{7-p}{5-p}}(bz)\right)
\ee
where in the above $J$ denotes the Bessel function, $N$ denotes the Neumann function. The near solution
$\tilde{g}_0(z)$ therefore takes the form,
\be\label{solnnear1}
\tilde{g}_0(z) = i z^{\frac{7-p}{5-p}} \left(J_{\frac{7-p}{5-p}}(bz) + i N_{\frac{7-p}{5-p}}(bz)\right)
\ee
Now the two solutions \eqref{solnfar1} and \eqref{solnnear1} can be matched in the common region when $\omega\rho_p \ll 1$ and this
determines the constant $c_\infty$ in terms of known constants as,
\be\label{constantreln}
c_\infty = \frac{1}{\pi} 2^{\frac{(7-p)^2}{2(5-p)}} b^{-\frac{7-p}{5-p}} \Gamma\left(\frac{7-p}{5-p}\right)\Gamma\left(\frac{9-p}{2}\right)
\ee
Therefore, we can write the solutions both in the near region ($u \ll 1$ or $(\omega\rho_p)^{\frac{7-p}{2}} \ll z \ll \omega\rho_p$ and
$z \gg b^{-\frac{5-p}{7-p}}(\omega\rho_p)$) and in the far region ($u \gg \omega\rho_p$) as follows,
\bea\label{finalsoln}
& & \tilde{g}_0(z) = i z^{\frac{7-p}{5-p}} \left(J_{\frac{7-p}{5-p}}(bz) + i N_{\frac{7-p}{5-p}}(bz)\right)\nn
& & \tilde{g}_{\infty}(u) = \frac{1}{\pi} 2^{\frac{(7-p)^2}{2(5-p)}} b^{-\frac{7-p}{5-p}} \Gamma\left(\frac{7-p}{5-p}\right)\Gamma\left(\frac{9-p}{2}\right)
u^{-\frac{7-p}{2}} J_{\frac{7-p}{2}}(u)
\eea 
Once we have the function $\tilde{g}(u)$, we can obtain the form of the graviton by using \eqref{tildegu} and \eqref{grav} as,
\be\label{grav1}
h_{\mu\nu} = \epsilon_{\mu\nu} \frac{\tilde{g}(\omega\rho)}{\sqrt{F(\omega\rho)G(\omega\rho)^{1-\frac{\d}{2}}}} e^{i\omega t}
\ee
The graviton flux can then be calculated from \eqref{grav1} using the standard definition as,
\be\label{flux1}
F = i\int_{\rho=\rho_S} \sqrt{-\bar g} e^{-2\bar\phi} \bar{g}^{\rho\rho}\left((\partial_\rho h_{\mu\nu}^{\ast}) h^{\mu\nu} 
- h_{\mu\nu}^{\ast}\partial_\rho h^{\mu\nu}\right)
d^{p+1}x d\Omega_{8-p} 
\ee
The integral is over a constant surface of radius $\rho=\rho_S$. We use \eqref{grav1} with the solutions \eqref{finalsoln} for the 
incoming waves of
the graviton in \eqref{flux1} to obtain the absorption cross-section as,
\bea\label{absorption}
\sigma_p (b,\omega,\rho_p) &=& \frac{(2\pi)^{8-p}}{\omega^{8-p} \Omega_{8-p}}\left|\frac{F_0^{\rm in}}{F_\infty^{\rm in}}\right|\nn
&=& \frac{\pi^{\frac{11-p}{2}}\left(\frac{2}{5-p}\right)^{\frac{9-p}{5-p}}}{2^{\frac{4}{5-p}} \left[\Gamma\left(\frac{7-p}{5-p}\right)\right]^2
\Gamma\left(\frac{9-p}{2}\right)}
b^{\frac{2(7-p)}{5-p}} \omega^{\frac{9-p}{5-p}} \rho_p^{\frac{(7-p)^2}{5-p}}
\eea  
where $b$ is a particular combination of parameters of non-susy D$p$ brane solutions defined in \eqref{defs} and 
$\Omega_d = \frac{2\pi^{\frac{d+1}{2}}}{\Gamma(\frac{d+1}{2})}$ is the volume 
of a $d$-dimensional unit sphre. $F^{\rm in}_0$ and $F^{\rm in}_\infty$ are the graviton fluxes for the incoming waves in the near and the far 
regions respectively. We thus see that the absorption cross-sections \eqref{absorption} for the non-susy D$p$ branes depend on all the three
parameters of the solutions, namely, $\rho_p$, $\theta$ and $\delta$ (through $b$) as expected. 
Now since $\rho_p \sim \ell_s$,  the absorption cross-sections in string units take the form,
\be\label{absorption1}
\frac{\sigma_p(b,\omega,\rho_p)}{\ell_s^{8-p}} \sim (\omega\rho_p)^{\frac{9-p}{5-p}},
\ee
which indeed vanish in the decoupling limit $\omega\rho_p \to 0$ as long as $p \leq 4$, showing that graviton does
not reach the branes or it decouples for non-susy D$p$ branes for $p \leq 4$. We will discuss $p=5,\,6$ cases
separately. The formula of the absorption cross-sections for the BPS D$p$ branes can be recovered from \eqref{absorption} by the BPS limit we 
discussed before. The BPS limit is given as $\rho_p \to 0$ and $\theta \to \infty$ (or $b \to \infty$) such that the product 
$b^2 \rho_p^{7-p} = \bar{\rho}_p^{7-p} =$ fixed. We note that in this limit $\sigma_p(b,\omega,\rho_p) \to \sigma^{\rm BPS}_p(\omega, \bar{\rho}_p)$,
where $\sigma_p^{\rm BPS}$ is the absorption cross-section on the BPS D$p$ branes obtained before in \cite{Alishahiha:2000qf}\footnote{We can 
compare the BPS 
results we obtain with those given in the Table 1 of \cite{Alishahiha:2000qf}. We find that there are some mismatch in the expressions of 
$\sigma_p$ for $p=1,2$. 
The numerical factors for $p=1$ we get is $\frac{\pi^4}{12}$ instead of $\frac{2\pi^4}{3}$ and for $p=2$ the factor we get is 
$\frac{\pi^2 (\Gamma(1/3))^2}{\sqrt[3]{3}\, 5}$ instead of $\frac{\pi^3 (\Gamma(1/3))^2}{\sqrt[3]{3}\, 2^2\, 5}$. For $p=3,4$ the numerical
factors match with our results.}.    

For $p=5$, the coordinate transformation in the near region \eqref{coordtrans} does not work and so the same method of obtaining the absorption 
cross-section would be problematic. Anyway, we will look at the potential \eqref{scatteringeq} and argue how the decoupling occurs in this case.
The potential \eqref{scatteringeq} for $p=5$ takes the form,
\be\label{potentialp5}
V(u) = \frac{3}{4u^2} - \frac{1}{4} \left(\frac{\partial_u G(u)}{G(u)}\right)^2 - F(u) G(u)^{-\frac{1}{2}}
\ee
We simplify the potential in the near region $\omega\rho_5 \ll u \ll 1$ (this is possible when $\omega\rho_5\ll 1$) as we have done for other 
D$p$ brane cases before. Then the potential \eqref{potentialp5} takes the form,
\be\label{potentialp51}
V(u) = \frac{3}{4u^2} - \left[1 + (b^2-\frac{1}{2})\frac{(\omega\rho_5)^2}{u^2}\right]
\ee
where $b$ is as defined in \eqref{defs}. If $b$ is ${\cal O}(1)$ or $\ll 1$, then the second term in the square bracket of \eqref{potentialp51} 
can be neglected as 
$u \gg \omega\rho_5$ and so, $V(u) = 3/(4u^2) - 1$. Also since $u \ll 1$, the first term dominates and eventually becomes infinite in the near
region. So, the graviton won't be able to reach the brane surmounting this infinite barrier and there is decoupling of gravity. On
the other hand when $\theta$ is such that $b\omega\rho_5 \gg u$, then the potential becomes,
\be\label{potentialp52}
V(u) = \frac{3}{4u^2} - b^2\frac{(\omega\rho_5)^2}{u^2}
\ee   
In this case, the potential acts as an infinite barrier as long as $\omega < \frac{\sqrt{3}}{2}\frac{1}{b\rho_5}$ and there is a decoupling
and if  $\omega > \frac{\sqrt{3}}{2}\frac{1}{b\rho_5}$, gravity couples to D5 brane. So, this gives a restriction on the energy of the incident
graviton for the decoupling to occur. This situation is very similar to the BPS D5 brane discussed in \cite{Alishahiha:2000qf}.

For $p=6$, the scattering potential \eqref{scatteringeq} takes the form
\be\label{potentialp6}
V(u) = - \frac{1}{4} \left(\frac{\partial_u G(u)}{G(u)}\right)^2 - F(u)
\ee
Now since both the terms here are negative, there is no maximum anywhere and it is a monotonically decreasing function as $u$ varies from
$\infty$ to $0$ without any barrier. Therefore, gravity will couple to non-susy D6 branes similar to BPS D6 branes.

We would like to remark that in obtaining the closed form expressions for the absorption cross-sections for the graviton we solved
the scattering equation \eqref{scatteringeq} analytically. However, in solving \eqref{scatteringeq} we had to assume two conditions 
(i) $\omega\rho_p \ll 1$ (needed to have an overlapping region for the far and the near brane solutions) and also (ii) $b$ (or $\theta$)
$\gg 1$ (needed to have a closed form solution of \eqref{scatteringnear}). Note that these two conditions\footnote{One might think that
these two conditions suggest that non-susy branes are becoming BPS branes in the decoupling limit and what we are obtaining is basically the
decoupling of gravity for these BPS branes and not the non-susy branes. However, the reason why this is not so is because these two
conditions are not correlated. Note that in order to get BPS branes we must take $b \to \infty$ and $\omega\rho_p \to 0$ simultaneously
such that the product $b^2 (\omega\rho_p)^{7-p} = (\omega\bar{\rho}_p)^{7-p}$ = finite. Here in our calculation, there is no relation
between the two conditions and they are taken independently. So, when we take the decoupling limit $\omega\rho_p \to 0$, $b$ remains 
large but finite and therefore, the non-susy branes remain non-susy.  But if we take $b$ also to infinity in the correlated way we 
just mentioned, then we recover graviton scattering cross-section result for the BPS brane as expected.}    
together imply that the non-susy
D$p$ brane solutions are near-BPS or near-extremal. So, our result for the absorption cross-sections \eqref{absorption} can be trusted
only in this near-extremal case. When we are far away from the extremality point there can be large corrections to the absorption 
cross-sections. But since our numerical results suggest that the decoupling must occur even when we are far away from the extremality 
point, the corrections must vanish in the decoupling limit $\omega\rho_p \to 0$, which implies that the corrections must involve a 
positive power of $\omega\rho_p$.       

\vspace{.5cm}

\noindent{\it 6. Conclusion} : To conclude, in this paper we have studied graviton scattering on the non-susy D$p$ branes of type II string
theories. We obtained the linearized equation of motion of the graviton in the background
of non-susy D$p$ branes. We have shown that both the minimally coupled scalar and the graviton essentially satisfy the same 
Schr\" odinger-like scattering equation
and from there we identify the potential the minimally coupled scalar or the graviton experiences while moving in this 
background. For all non-susy D$p$ brane 
backgrounds we found that far away from the branes the potentials go over to $-1$, while near the branes they take the value $-\infty$ as in
the case of BPS D$p$ branes. However, because of the complicated form of the potentials it is difficult to understand its behavior in between.
So, we have studied them numerically and plotted the potentials $V(u)$ versus $u$ in Figures 1, 2, 3 for various values of $p$ in each Figure. As 
the non-susy branes have three independent 
parameters $\rho_p$, $\theta$ and $\delta$, the potentials also depend on them. For simplicity we have set $\theta=0$ and $\delta=\delta_c/2$, where
$\delta_c$ is the maximum allowed value of $\delta$ described in section 2 and then plotted $V(u)$ for three different values of $\omega\rho_p$
(where $\omega$ is the energy of the graviton), namely, 1.0, 0.1 and 0.01 in the three Figures mentioned above to show the behavior in the 
intermediate region. We found that there are maxima in the potentials near the origin for each value of $p\leq 5$, but the maximum is absent
for $p=6$. When we lower the value of $\omega \rho_p$, the same feature remains, but the height of the maxima rise sharply and they shift 
more towards the origin. We thus concluded that when we take the decoupling limit $\omega \rho_p \to 0$, there will be infinite barriers close to
the origin for all $p \leq 5$, but for $p=6$, there will be no barrier and the potential will decrease monotonically to $-\infty$. Thus the graviton
will not be able to reach the brane for $p \leq 5$ and there will be decoupling of bulk gravity from the brane world volume. On the other hand,
for $p=6$, since there is no barrier, graviton will couple to the brane.

To give further support to our claim, we tried to solve the graviton scattering equation for the non-susy D$p$ branes. We found that we can
write a closed form solution in the far region away from the branes, but it is difficult to solve the differential equation in the near region
for all $u$. We found that a closed form solution in the near region is possible if we go to a certain range of space within the near region. 
To make this possible we found that the parameters of the solutions must satisfy the conditions (i) $\omega\rho_p \ll 1$ and (ii) 
$b({\rm or}\,\, \theta)\gg 1$. We have emphasized that this does not mean that decoupling occurs only for large values of $b$. Decoupling 
occurs even for
small values of $b$, but the closed form solutions is possible only if we use this condition. From these solutions we have computed the graviton
absorption cross sections for the non-susy D$p$ branes and found that they depend on the parameter $b$ as well as some positive powers of 
$\omega \rho_p$ for $p\leq 4$. Thus we found that in the decoupling limit $\omega\rho_p \to 0$, the graviton absorption cross-sections
vanish for $p \leq 4$ and therefore gravity decouples. We have also compared
our results with the BPS results obtained before in \cite{Klebanov:1997kc, Gubser:1997yh, Alishahiha:2000qf}. For $p=5$, we were not able
to solve the equation even with the conditions mentioned above, but by analyzing the scattering potential we concluded that in this case if the 
parameter $b$ is ${\cal O}(1)$ or $\ll 1$, then gravity decouples for all energies of the graviton (unlike the BPS D5 brane), but if $b \gg 1$, 
then gravity decouples only if the energy of the graviton is below certain critical value, otherwise it couples, similar to BPS D5 brane.           
Even for $p=6$, we argued by analyzing
the scattering potential that the gravity in this case couples to D6 brane as there is no barrier in the potential for the graviton.
Thus, contrary to what is known in the literature \cite{Brax:2000cf, Brax:2001dc}, we have shown here that there is a decoupling of gravity 
on the non-susy D$p$ branes very similar to the BPS branes, modulo some differences for the case of D5 brane. We remarked that though
the decoupling of gravity occurs in general for the non-susy D$p$ branes the closed form expressions for the 
absorption cross-sections can only be obtained for the near-BPS or near-extremal point. The expressions derived in \eqref{absorption}
can not be trusted far away from the extremality and there can be large corrections proportional to some positive power of $\omega\rho_p$.      

So far we have not discussed anything about the boundary theory or the theory on the brane. It would be very interesting to understand
this aspect in detail. However, we would like to mention that since we are dealing with non-susy branes, there may be open string tachyon
\cite{Sen:1999mg, Sen:2004nf}
living on the brane. When $\theta =0$, the charge of the non-susy branes vanish (see \eqref{relations}) and the world volume theory in 
this case would be purely tachyonic field theory. But when $\theta \neq 0$, there are gauge fields on the brane and so, the theory in this
case would be pure Yang-Mills theory coupled with tachyon. It can be checked that when $\alpha+\beta=2$ and $\theta$ is large, the gravity
theories reduce to some deformations of the near horizon geometry of BPS D$p$ branes. Thus the world volume theories would be some 
perturbations of pure Yang-Mills theories in various dimensions without any tachyon field. We hope to come back on these issues along 
with others in future.

\vspace{.5cm}

\noindent{\it Acknowledgements:} {\small We would like to thank the anonymous referee for raising some questions which made us realize that some
derivations we had given in the earlier version of this paper were not quite correct. We have made those corrections in this version although
that does not affect the main results, the graviton scattering, of the paper.}

\end{document}